\documentclass[aps,twocolumn,preprintnumbers,floatfix]{revtex4}
\usepackage{graphicx}
\usepackage{subfigure}
\usepackage{multirow}
\usepackage{float}
\usepackage{fancyhdr}
\usepackage{longtable}
\usepackage{parskip}
\usepackage[T1]{fontenc}
\usepackage{dcolumn}
\usepackage{siunitx}
\usepackage{bm}
\usepackage{amsfonts}
\usepackage{amsmath}
\usepackage{amssymb}
\usepackage{array}
\usepackage[bookmarks=false]{hyperref}

\newcommand{\pwisein}{\left\{ \begin{array}{ll}}
\newcommand{\pwiseout}{\end{array}\right.}

\begin{document}

\title{Thermal Capacity Mapping of Cryogenic Platforms for Quantum Computers}

\author{Scott A. Manifold, George B. Long and Jonathan J. Burnett}

\affiliation {\it Oxford Quantum Circuits, Reading, RG2 9LH}

\date{March 13, 2025}

\begin{abstract} 
Large-scale cryogenic Input-Output (IO) infrastructure is a requirement for realising fault-tolerant quantum computing in solid-state modalities. Such IO scaling presents significant challenges in thermal modelling, hardware design and verification. Here we present a design tool for cryogenic platform development with applicability to quantum computing and other cryogenic technologies. By taking comprehensive measurements of a commercially available dilution refrigerator we construct a ‘platform capacity map', quantifying the platform's complex response to heat loads distributed over multiple temperature stages and accounting for interstage dependencies, avoiding an analytical model. This allows for the modification of the platform parameters to be accurately estimated for the inclusion of an arbitrary IO infrastructure, the identification of stage thermal overheads and the optimisation of system parameters. We identify a key bottleneck around the Still stage, owing to loading from connected stages, and observe discrepancies between existing thermal modelling methods for estimating heat loads associated with IO and experimental results obtained from the platform capacity map. 
\end{abstract}

\maketitle 

\section{Introduction} \label{Introduction}
The development of computationally useful and eventually fault-tolerant quantum computers comes with considerable challenges, including scaling quantum processors to potentially millions of physical qubits \cite{Gidney2021howtofactorbit}. With regard to solid-state qubit platforms, this leads to the requirement of a cryogenic Input-Output (IO) infrastructure significantly larger than those in use today. This scaling requirement has driven efforts in alternative IO approaches, including moving control components to cryogenic temperatures. This serves to decrease both the number of interconnects and potentially the control latency of an error correction loop, with examples of prerequisite technologies including: Si CMOS \cite{PhysRevApplied.3.024010}, high electron mobility transistors (HEMT) \cite{Ferraris2024}, SiGe BiCMOS \cite{4838951}, photonic links \cite{Lecocq2021}, optical transduction \cite{Mirhosseini2020} and single flux quantum (SFQ) logic \cite{80745}. Regardless of the exact implementation used to provide a scaled cryogenic infrastructure for future large quantum processors, understanding the performance of cryogenic platforms and thermal overhead is of critical importance for practical realisation. However, the tooling around thermal modelling of cryogenic IO infrastructure is generally limited to a component-level view of how much heat is dissipated within a static model of an isolated temperature stage of a cryogenic platform, or with passive heat loads due to stage connections calculated via approximate models \cite{Krinner2019}, with both reliant on the availability of material reference data over the relevant temperature ranges. 

In this work a novel approach to thermally modelling cryogenic platforms and the inclusion of cryogenic infrastructure is presented as a design tool for scaling cryogenic IO. These techniques are relevant to base cryogenic system design, quantum computing systems and other cryogenic devices including superconducting single-photon detectors \cite{Oripov2023} for photonics applications. We find that when designing IO platforms compatible with large quantum processors or scaled cryogenic infrastructure, the commonly held perception of not exceeding an independent heat load budget per stage is insufficient. Here we report on an alternative approach that uses an empirical platform thermal capacity map to accurately describe the response of the platform to an arbitrary distributed heat load and inform on system-level thermal overheads. Furthermore, the platform capacity map can be combined with thermal modelling of the IO payload to allow for the estimation of payload compatibility with the platform and modification to platform parameters (stage temperatures and circulation pressures). 

\section{Platform capacity mapping} \label{btlcm}
\begin{figure*}
\includegraphics[width=7in]{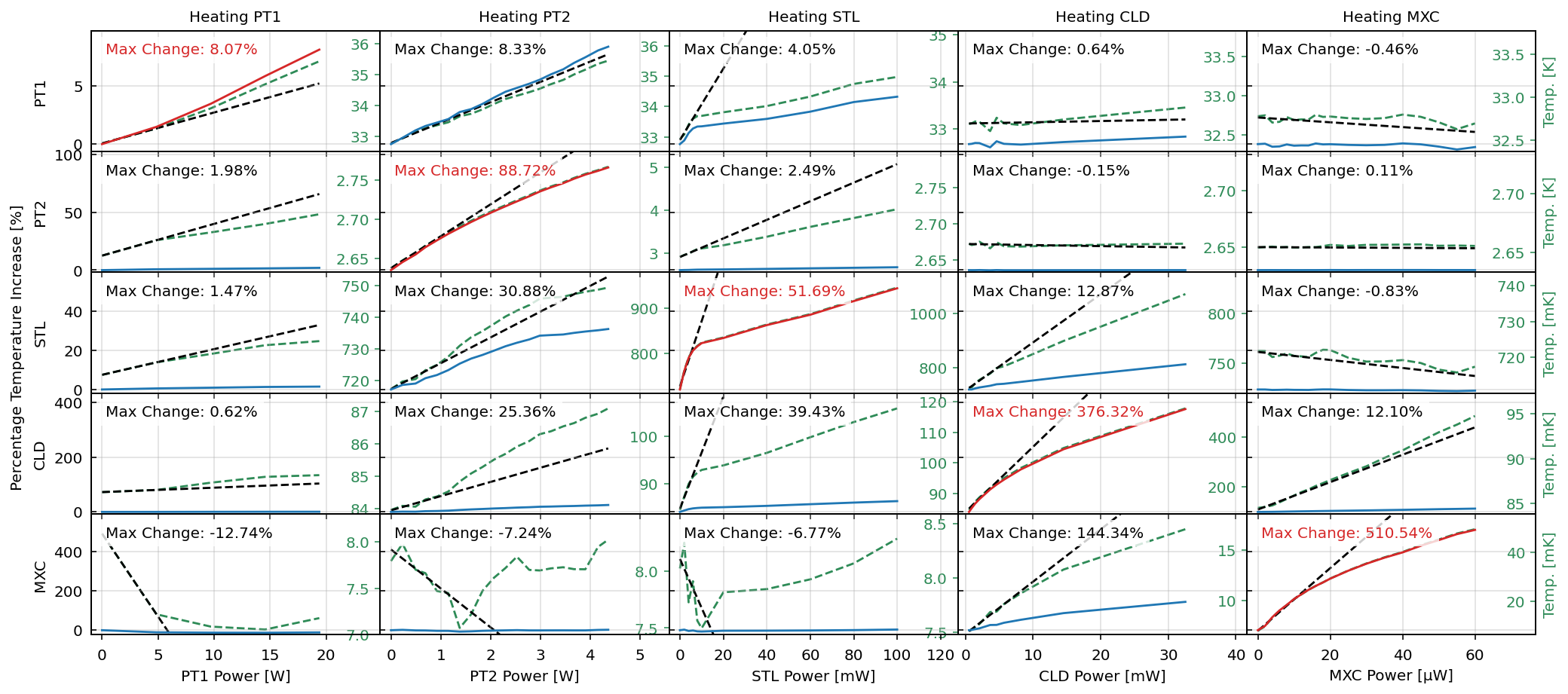}
\caption{\label{staticshot} 
Red and blue solid lines show stage percentage temperature change as a function of loading any individual stage. Red lines indicate the temperature response of the stage being heated while blue lines show the response of other stages. Green dashed lines show the corresponding absolute temperature response and dashed black lines represent predictions of the linear thermal model.
}
\end{figure*}

A typical approach used to model the impact of the cryogenic infrastructure on the temperature stages of a dilution refrigerator is to use a linear approximation where it is assumed that a temperature change of each stage ($\Delta T$) is a function of the thermal loads applied at all stages. This can be represented by a system of linear equations derived from the matrices:
\begin{eqnarray}
\Delta T = A \cdot Q
\label{eq:five},
\end{eqnarray}
\begin{eqnarray}
\begin{bmatrix}
\Delta t_{11} & \cdots & \Delta t_{1j} \\
\vdots & \ddots & \vdots \\
\Delta t_{i1} & \cdots & \Delta t_{ij} \\
\end{bmatrix}
\label{eq:six}
=\begin{bmatrix}
a_{11} & \cdots & a_{1j} \\
\vdots & \ddots & \vdots \\
a_{i1} & \cdots & a_{ij} \\
\end{bmatrix} \cdot
\begin{bmatrix}
q_1 \\
\vdots \\
q_{j}\\
\end{bmatrix}
\end{eqnarray}
where $\Delta T$ contains the temperature changes $\Delta t_{ij}$ of measurement stage $i$ when loading stage $j$, $A$ is the coupling matrix representing all coefficients $a_{ij}$ describing the thermal coupling between the stages and $Q$ is the vector describing the thermal load applied per stage $q_j$, referred to as the loading scenario. The cooling power for stage $j$ can be stated at temperatures corresponding to change $\Delta t_{jj}$ given $q_j$.

This approach relies on the assumption that temperature changes $\Delta T$ for each stage of the dilution refrigerator: first stage pulse tube (PT1), second stage pulse tube (PT2), Still (STL), cold plate (CLD), and mixing chamber stage (MXC) are linearly related to the thermal load $Q$ applied to every other stage through the coupling matrix $A$. We compare this method with measurements of an as delivered \textit{Oxford Instruments Proteox LX 850}, by sequentially applying a varying heat load to each individual stage and recording the modification to all stage temperatures. A description of the platform specification and measurement methodology used throughout this study is given in Suppl. \ref{msetup} and Suppl. \ref{method} respectively. 

The temperature response of each stage to individual stage loading is shown in Fig. \ref{staticshot}. The linear approximation (Eq. \ref{eq:six}) is only an accurate representation of stage temperature when the heat load applied to any individual stage is sufficiently small (significantly less than the apparent maximum available cooling power on each stage); as the heat load is increased, non-linear behaviour is observed which leads to a growing disagreement. This demonstrates that even in the simplest situation, where the thermal loading scenario is restricted to only a single stage rather than a more realistic and complex loading scenario involving the application of heat to multiple stages, the linear approximation holds only within a limited range, implying a false design ceiling on payload IO that may result in an over or underestimation. 

\begin{figure}[t]
\includegraphics[width=3.3in]{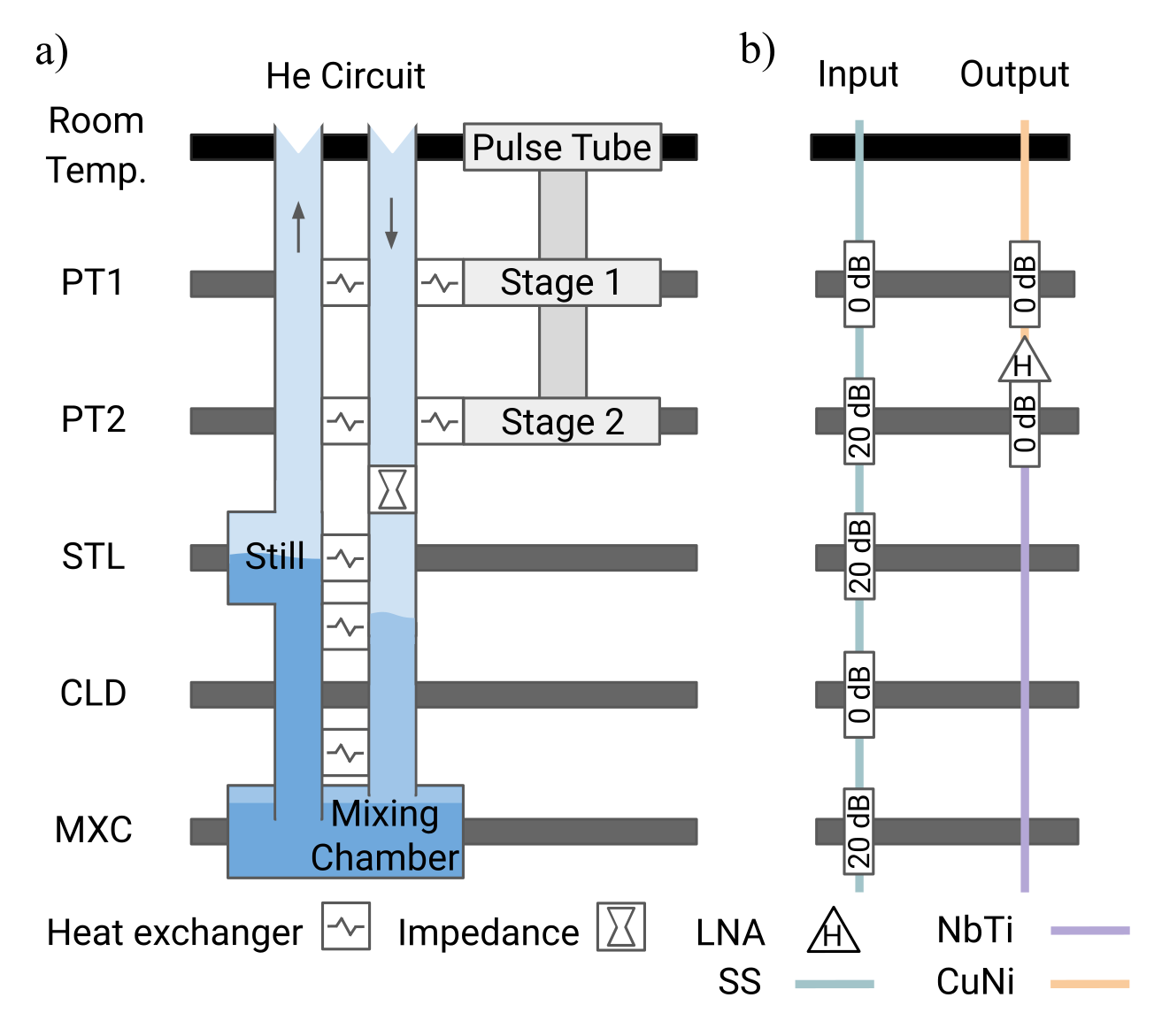}
\caption{\label{DR Cart} 
a) Simplified schematic of a dilution refrigerator, showing key components of the $^3$He/$^4$He circulation circuit. b) Example of a RF IO channel configuration typically used for control and readout of superconducting qubits, where low thermal conductivity stainless steel (SS) highly attenuated transmission lines provide input signal connectivity for control and readout pulses, with readouts signals routed out of the cryostat amplified by a HEMT cryogenic low noise amplifier (LNA), using superconducting niobium titanium (NbTi) and lower attenuation copper nickel (CuNi) transmission lines.}
\end{figure}

\begin{figure*}
\includegraphics[width=6.7in]{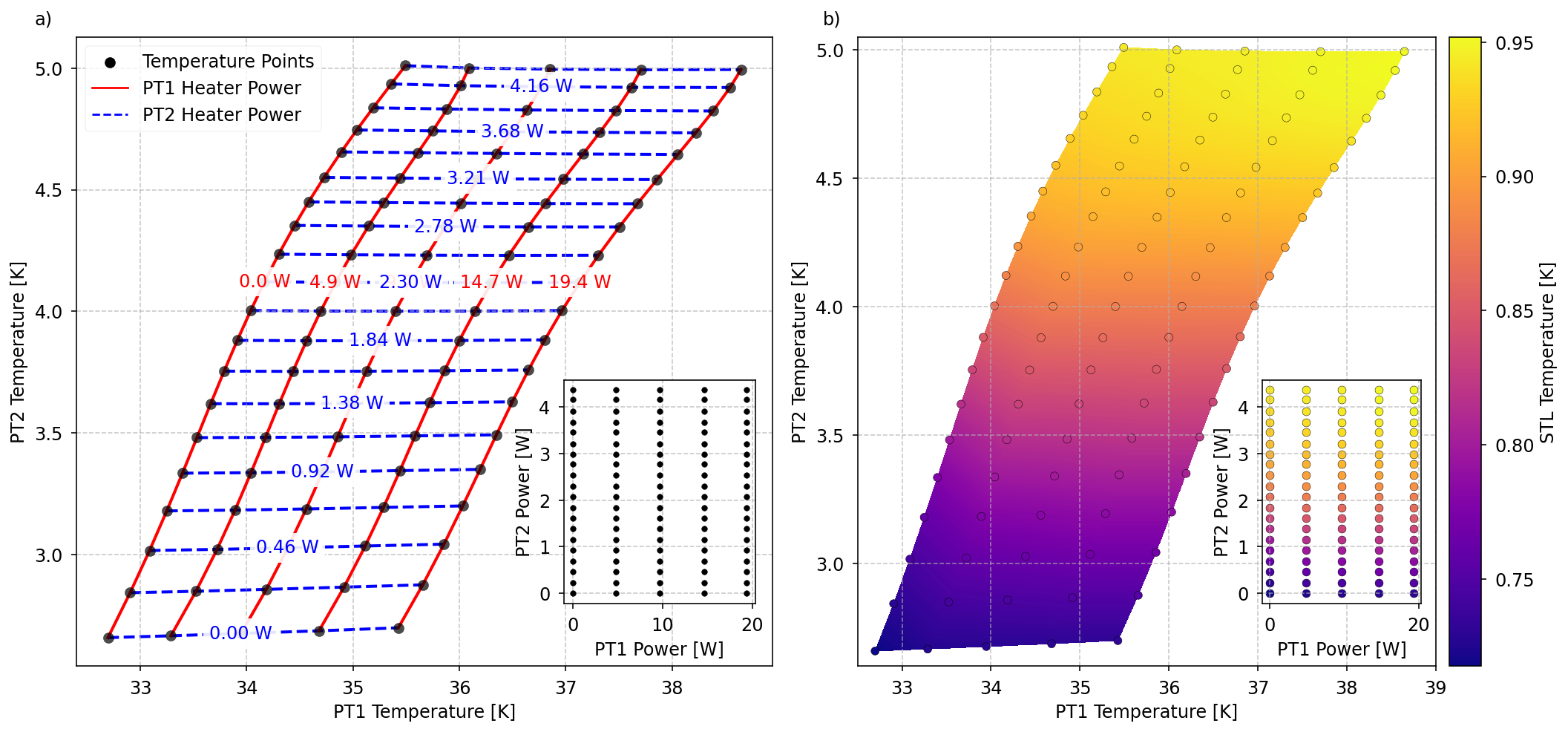}
\caption{\label{capmap} 
Pulse-Tube (PT) map. PT1 and PT2 temperatures are plotted as a function of power applied to the PT1 and PT2 stage, in this instance \SI{0}{\watt} is applied to the STL, CLD and MXC heaters. The power space is plotted in the inserts, ranging from \qtyrange[range-units = single]{0}{4.26}{\watt} and \qtyrange[range-units = single]{0}{19.4}{\watt} on the PT2 and PT1 stage respectively. a) Red/Blue lines and annotations indicate PT1/PT2 power. b) The measured STL temperature over the PT2 and PT1 power range is interpolated and displayed as a colour map. 
}
\end{figure*}

These limitations arise in-part from the simple treatment of the dilution refrigerator and the complex relationships governing individual-stage cooling power and interstage dependencies. An illustration of a dilution refrigerator and a typical He circuit is shown in Fig. \ref{DR Cart}a). The \textit{Proteox LX 850} used in this study, combines pulse tube cryocoolers with an inline impedance and heat exchangers to pre-cool and condense a mixture of $^3$He and $^4$He gas to achieve separation into two distinct liquid phases (concentrated and dilute) with a phase boundary located in the MXC. The flow of liquid He through the heat exchangers and mixing chamber is complex, with, for example, Kapitza thermal resistance and viscosity not only creating strong temperature dependence but also a sensitivity to the location of the phase boundary \cite{pobell2007matter}. Cooling originates from $^3$He atoms crossing this phase boundary, with a rate controlled by heat load on the STL, with the outgoing $^3$He re-condensed (circulated) using a series of pumps to allow for continuous operation of the system at milli-Kelvin temperatures. The rate of evaporation of $^3$He in the STL, characterised by the flow rate through the STL pumping line, is determined by the STL temperature. The STL temperature is related to factors including: pulse tube stage temperatures, He circulation circuit, heat exchangers and active or passive heat loads present on the STL itself. These considerations create a dynamic system that cannot be easily captured by an analytical model without an extensive understanding of the system's individual components.

To overcome this, we present an alternative method for accurately determining the response of the system to thermal loading by constructing a platform thermal capacity map to quantify the impact of a distributed heat load on the refrigerator. Mapping is performed in the absence of a payload, capturing all intrinsic heat loads including conductive, convective and radiative, inherent to the system and assumed not to be modified by the addition of cryogenic infrastructure. The platform map is obtained by systematically varying the heat applied to each individual stage as a function of the heat applied to every other stage to form a coordinate system of stage thermal loads (PT1, PT2, STL, CLD and MXC). Details of the power parameter space can be found in Suppl. \ref{method}. For each coordinate in this space, platform parameters (inclusive of stage temperatures, circulation pressures, and helium flow rate) are recorded. Thermal capacity maps, with a basis of a two-dimensional (2D) heater power grid, are commonly used to describe the relationship between stage temperature and cooling power of two-stage Gifford-McMahon (GM) and Pulse Tube (PT) cryocoolers \cite{cryomech}. To the best of our knowledge, this is the first time this concept has been reported for an entire dilution refrigerator.

The platform capacity map is harder to visualise than a traditional 2D thermal capacity map used for two-stage cryocoolers given the extra dimensions introduced by the additional stages of the platform and increased amount of measured parameters. As a starting point, it is instructive to define two important slices of the platform capacity map, a Pulse-Tube (PT) map and a Dilution Unit (DU) map. 

A PT map, shown in Fig. \ref{capmap}a), is equivalent to a traditional 2D thermal capacity map for a GM/PT cryocooler, with the PT1 and PT2 stage temperature recorded as a function of the PT1 and PT2 power with the heat load applied to other stages (STL, CLD, MXC) held constant. The \textit{Proteox LX 850} has two \textit{Cryomech PT420-RM} pulse-tube cryocoolers providing cooling power at the PT1 and PT2 stages. The \textit{PT420-RM} specification indicates a cooling power of approximately \SI{1.8}{\watt} at \SI{4.4}{\kelvin} for PT2 with \SI{25}{\watt} at \SI{33.5}{\kelvin} for PT1 for each unit. However, the PT map shown in Fig. \ref{capmap}A indicates an available cooling power of approximately \SI{2.78}{\watt} and \SI{4.9}{\watt} at \SI{4.4}{\kelvin} and \SI{36}{\kelvin} respectively. Assuming it is valid to double the manufacturer specified cooling power at these temperatures to account for presence of two identical units, this corresponds to a reduction of \SI{0.82}{\watt} at PT2 and \SI{45.1}{\watt} at PT1, which can be primarily attributed to additional background heat loads due to integration with the dilution refrigerator and the dynamics of He circulation. Fig. \ref{capmap}b) shows the same PT map now with the STL temperature superimposed. Contrasting to Fig. \ref{staticshot} the effect on the STL temperature from loading multiple stages can now be understood and accurately quantified for various scenarios of distributed heat load applied to the PT1 and PT2 stages.

\begin{figure}[ht]
\includegraphics[width=3.3in]{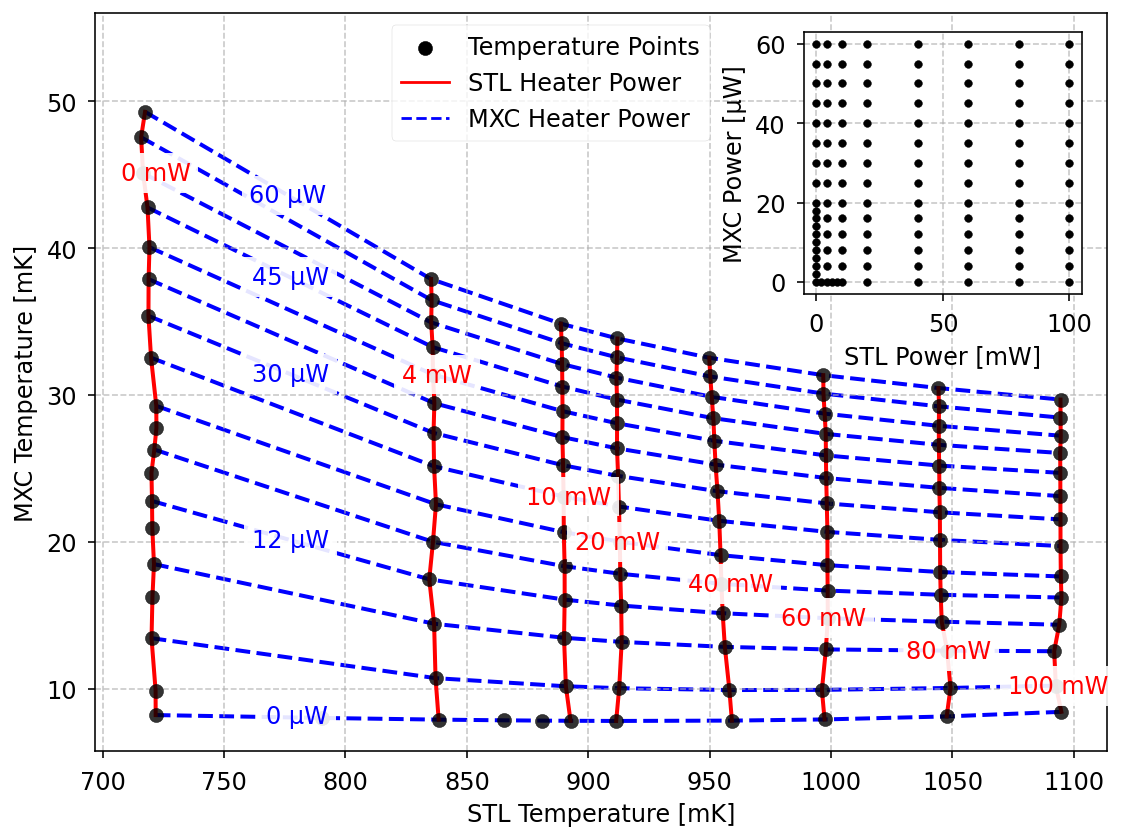}
\caption{\label{slicemap} 
Dilution Unit (DU) map. MXC and STL temperatures are plotted as a function of power applied to the MXC and STL stage, in this instance \SI{0}{\watt} applied to PT1, PT2 and CLD heaters. The power space is plotted in the insert, ranging from \qtyrange[range-units = single]{0}{60}{\micro\watt} and \qtyrange[range-units = single]{0}{100}{\milli\watt} on the MXC and STL stage respectively. Red/Blue lines and annotations indicate Still/MXC power.
}
\end{figure}

A DU map, shown in Fig. \ref{slicemap}, is similar to the PT map however, now the STL and MXC temperature are plotted as a function of the STL and MXC power with the heat applied to the other stages (PT1, PT2, CLD) held constant. The DU map represents a useful way of understanding and visualising the non-linear relationship between STL temperature and available cooling power at the MXC stage, of critical importance for optimising cryogenic infrastructure design. The modification of the DU map for various loading scenarios (PT2 and CLD loading) are discussed in Section \ref{lpm} providing insight into the thermal overhead of the platform.

PT and DU maps allow for a simple way to visualise the platform's response to various loading scenarios as well as the effect of their loading on other stages and modification of platform parameters. The platform capacity map can be sliced in many different ways depending on which relationship between stages and loading scenario is of interest. Although analysing the platform map is a constructive exercise when considering cryogenic IO design, it can also be used as a powerful tool to accurately model the inclusion of the distributed heat load over the full operational range of the platform. 
\begin{table*}[t]
\begin{tabular}{c|cc|cc|}
\cline{2-5}
\multicolumn{1}{l|}{} & \multicolumn{2}{c|}{Measured} & \multicolumn{2}{c|}{Prediction}      \\ \hline
\multicolumn{1}{|l|}{Stage} & \multicolumn{1}{c|}{Temperature} & Inferred Heat Load & \multicolumn{1}{c|}{Inferred Temperature} & Estimated Heat Load \\ \hline
\multicolumn{1}{|c|}{PT1}   & \multicolumn{1}{c|}{33.14 K}     & 2.63 W    & \multicolumn{1}{c|}{33.02-33.36 K}     & 1.99-4.45 W   \\ \hline
\multicolumn{1}{|c|}{PT2}   & \multicolumn{1}{c|}{2.724 K}     & 76.94 mW  & \multicolumn{1}{c|}{2.69-2.75 K}      & 35.2-102.07 mW \\ \hline
\multicolumn{1}{|c|}{STL}   & \multicolumn{1}{c|}{739.5 mK}    & 394.81 µW & \multicolumn{1}{c|}{729.37-739.02 mK}   & 100.14-218.66 µW  \\ \hline
\multicolumn{1}{|c|}{CLD}   & \multicolumn{1}{c|}{92.55 mK}    & 243.44 µW & \multicolumn{1}{c|}{85.25-86.66 mK}    & 6.28-16.82 µW   \\ \hline
\multicolumn{1}{|c|}{MXC}   & \multicolumn{1}{c|}{8.71 mK}    & 1.82 µW   & \multicolumn{1}{c|}{7.97-10.43 mK}     & 0.06-4.51 µW    \\ \hline
\end{tabular}
\caption{\label{vladtable} Prediction vs measured results for the validation payload. The LNA compliment is unpowered for this test. The measured stage temperatures are input into the platform capacity map to give the inferred heat load introduced by the payload. The estimated heat load range corresponds to the minimum and maximum modelled heat load per stage obtained by using different material data references and modelling methods for the validation payload as discussed in Suppl. \ref{payloadcomp}. The minimum and maximum estimated heat load are input into the capacity map to give a corresponding range of inferred temperature predictions.}
\end{table*}

\section{Payload Modelling}\label{pmodel}
Application specific payloads will lead to variation in IO channel density, components and applied biases. However, regardless of how the payload is populated, it invariably constitutes a combination of passive and active sources of heat loading to each stage of the platform. By calculating the distributed heat load presented by a prospective payload we can input this loading scenario directly into the capacity map acquired in Section \ref{btlcm}. This allows for a predictive modelling capability, allowing for an estimate of the modification to the system parameters for any arbitrary payload that fits within the power space covered in the platform capacity map. 

To demonstrate this, a validation payload is installed in the same cryostat the platform capacity mapping measurements were performed. The validation payload incorporates an IO channel configuration, illustrated in Fig. \ref{DR Cart}b), populated with cryogenic radio-frequency (RF) componentry typical for superconducting qubit measurement \cite{Krinner2019}. The payload consists of 64 highly attenuated coaxial input lines and 24 amplified coaxial output lines. A detailed description of the validation payload and the method used to calculate the heat load presented by its componentry is detailed in Suppl. \ref{payloadcomp}. 

Post cool-down, a coarse version of the platform capacity mapping is performed to establish that the platform's response to loading is consistent with the previous measurements with the exception of an offset on each stage corresponding to heat load introduced by the validation payload. Care and standard operating protocols were enacted to reduce the potential for error from cooldown-to-cooldown variance, discussed in Suppl. \ref{method}, and we find that the platform capacity map is consistent with the previous measurements. As such, the stage offsets can be attributed to an inferred heat load owing to the payload. The measured stage temperatures and inferred heat load can be compared with the results of the payload modelling where an estimated distributed heat load is input to the capacity map to infer the predicted stage temperatures, with the results summarised in Table \ref{vladtable}. 

\begin{figure*}[t]
\includegraphics[width=7in]{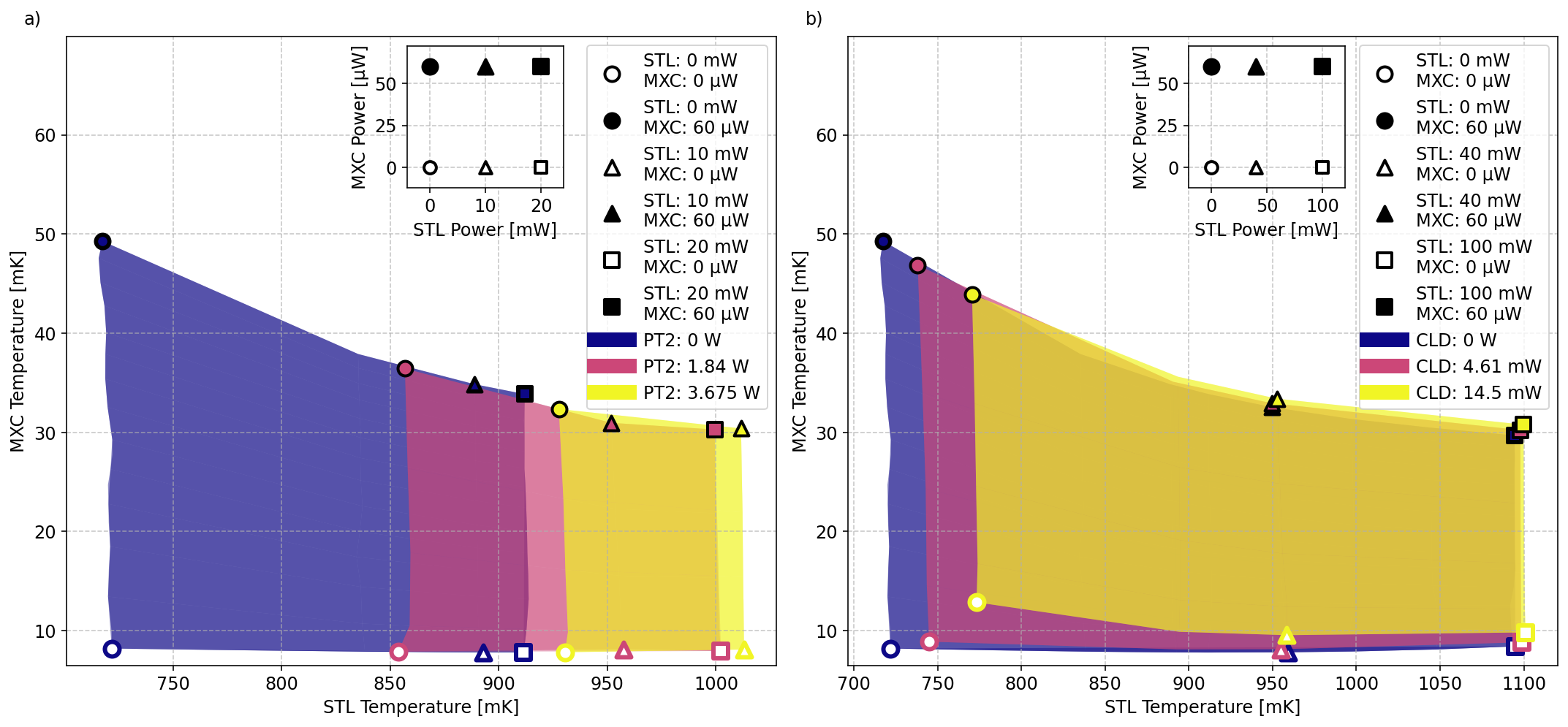}
\caption{\label{overheads} Individual DU maps overlaid for a) PT2 heater power of 0 W (blue), 1.84 W (purple) and 3.675 W (yellow), with the PT1 and CLD heater set to 0 W b) CLD heater power at 0 W (blue), 4.61 mW (purple) and 14.5 mW (yellow), with the PT1 and PT2 heater set to 0 W. Inserts highlights MXC and STL power points common to each map however the corresponding temperatures shifted due to the constant loading of PT2(a)/CLD(b) stage. The effect of the loading the CLD or PT2 stage on the STL/MXC overhead can be simply interpreted as the area of the capacity map shrinking representing a constraint to loading on the STL/MXC stages that can be afforded to the IO while staying within the specified limits.
}
\end{figure*}

The measured stage temperatures fall within or close-to the minimum and maximum inferred temperature range predictions. However, given the requirement of the IO payload to considerably increase in scale for future platforms it is important to note the disagreement between the estimated and inferred heat load, particularly on the STL and CLD stage, where these underestimations could significantly hamper predictions of compatibility with larger payloads. There are several factors that could lead to discrepancies in the calculation of the estimated heat loads. First, estimation of the temperature-dependant thermal conductivity of the cabling (Cu-Ni, NbTi and SS). As noted in Suppl. \ref{payloadcomp}, a significant difference in thermal conductivity of the cabling is found when comparing cable manufacturer specification with explicit modelling of the cabling using available references of thermal conductivity for the constituent materials, in both cases requiring extrapolation to sub-Kelvin temperature due to a lack of empirical data in this range. Second, assumptions around thermalisation. The SS and CuNi cabling is thermalised at each stage via a thermally anchored attenuator acting as a stage feed-through, with the predictions assuming a complete thermalisaton of both the inner and outer conductor. The NbTi cabling is continuous between the PT2 and MXC chamber stages. Although the outer conductor is mechanically clamped between stages the thermalisation of the inner conductor is less well constrained and how it thermalises is ambiguous. Furthermore, temperature dependant thermal contact resistances are also neglected, which could also contribute to incomplete thermalisation of the cabling at lower temperatures. Third, mechanical tolerances and material composition. For the predictions the situation is idealised to each cable being identical in construction, composition and mechanical fixing to the platform. Ultimately, there are numerous assumptions that are required to model the heat load associated with a realistic IO payload, in excess of the three listed here. In the preparation of this manuscript we became aware of recent work \cite{raicu2025} which extends understanding of interstage passive heat load models and reports findings broadly consistent with our own when modelling cables and with the use of similar approximations and assumptions. The method of measuring the inferred heat load using the platform capacity map outlined here avoids the assumptions complicating payload modelling and reliance on material reference data. As such, using this method to measure and validate associated payload heat loads provides a more accurate basis for large scale payload modelling and design.

\section{Discussion}\label{lpm}
Scaling challenges in cryogenic IO and the drive for alternative IO technologies in-part relate to circumventing thermal bottlenecks present on today's cryogenic platforms. However, typically these bottlenecks are isolated to a maximum permissible heat load threshold on each individual stage of the platform. Understanding the thermal overhead of the platform is critical to designing robust cryogenic platform that supports payload scaling. This requires a more holistic understanding of the platform and how its performance is modified by a distributed payload. 

In order to define an overhead we must first define limits of operation. For a quantum processor, thermal excitations results in noise detrimental to computation and a practical limit on a MXC temperature below 30 mK can be specified, for technologies operating around 5-10 GHz, to ensure the thermal photon number is sufficiently minimised. Similar limits can apply to additional stages where an operational temperature range must be provided to meet payload componentry requirements. However, not all limits manifest as an operational temperature requirement. When performing the platform capacity mapping measurements detailed in Section \ref{btlcm}, it was found that exceeding a maximum threshold in circulation pressure was frequently encountered even in situations where the MXC remained at an acceptable temperature. This was primarily encountered when the loading scenario warmed the STL temperature to above 1.1 K, resulting in an elevated STL line pressure and the turbo molecular pumps, responsible for $^3$He circulation, reaching their maximum power rating to maintain their optimal rotation speed. This illustrates the, often unstated, requirement that the platform He circuit and pumping configuration are suitable to support heavy loading scenarios that potentially come with a scaled IO infrastructure. The response of the circulation pressures to different loading scenarios is further expanded on in Suppl. \ref{pressureandflow}. Another typically overlooked limit is the operational up-time. With future large-scale quantum computing systems potentially running quantum programs with physical run times measured in months to years \cite{2211.07629}, the cryogenic platform must meet this resource requirement. 

To visualise overheads and how a distributed heat load impacts the available cooling power on any individual stage, it is constructive to look at how the DU or PT maps, defined in Section \ref{btlcm}, are modified for different loading scenarios. In Fig. \ref{overheads}a) three DU maps are superimposed for different PT2 loading scenarios (0 W, 1.84 W and 3.68 W applied to the PT2 heater). The effect of loading PT2 directly translates to a higher STL temperature for any given STL power, in turn increasing the flow of $^3$He in the circulation unit and allowing a greater power handling at the MXC stage. However, this also has the effect of greatly reducing the STL power overhead from 100 mW to 22 mW to 10 mW for a PT2 load of 0, 1.84, 3.675 W respectively. This reduction in the overhead corresponds to a reduction in the heat budget for the payload IO at the STL stage. This is further complicated when considering additional stages such as PT1 and the CLD stage, not shown in Fig. \ref{overheads}a) for clarity. The STL temperature is one of the most important parameters to effectively design around as it directly translates to MXC temperature and is typically critical to device (quantum processor or otherwise) performance. Fig. \ref{overheads}a) illustrates that the view that assigning a single threshold to the power dissipated at the STL stage is insufficient and this threshold must consider the addition of loading on all other stages.

Fig \ref{overheads}b) shows a different scenario where 4.61 mW and 14.5 mW is applied to the CLD stage. In this case the overheads are only slightly modified however, the minimum attainable temperature of the mixing chamber in the presence of the CLD load is increased for the same range of applied STL power. Although CLD loading does have an effect on the MXC temperature, the MXC remains at a generally acceptable temperature even with a substantial load applied to the CLD. The CLD is often seen as a key thermal bottleneck due to the lack of active cooling on this stage (only coming from coupling to the dilution unit heat exchanger), however, we find 3.7 mW of cooling power at 200 mK at the CLD stage, considerably larger that that seen in other measurements of this parameter \cite{Krinner2019}, \cite{POOLE2022103538}. This could stem from the lack of wide reporting on this particular overhead and differences in performance of commercially available dilution refrigerators. However, it represents a surprising result only obtained by performing this detailed characterisation of our platform, with the larger than expected overhead significant when considering IO payload design in this instance.

\section{Conclusion}
The scaling challenges posed by quantum computing and cryogenic technologies are numerous, with the increasing complexity and associated cost of the cryogenic infrastructure of key concern. This work highlights the benefit of a deeper understanding of cryogenic platforms and the utility of a platform capacity map for insights on system-level overheads and tooling towards modelling of scaled cryogenic IO infrastructure. Rather than being forced to define heat load budgets relative to an estimated cooling power corresponding only to maximum temperatures, platform mapping provides the certainty of establishing maximum distributed heat loads for any combination of stage temperatures, pressures or the flow rate. Furthermore, this method provides a framework for validating and predicting platform compatibility with prospective payloads without solely relying on an analytical passive heat load model and material reference data. 

The capacity mapping technique should find ready application within industry with cryogenic OEMs. Furthermore, it expands the toolset for manufacturers of components, IO, and quantum computing system designers to reliably design large-scale IO platforms, given that this approach allows for design confidence and the ability to optimise both the cryogenic platform and infrastructure in advance of manufacture and integration. A key extension of this work would be understanding consistency of platform capacity maps for different cryogenic systems to understand platform-to-platform performance variation as further validation of this approaches accuracy and applicability to a variety of cryogenic technologies. 

\section{Acknowledgments}
We extend our thanks to the OQC team and those who supported various aspects of this work. The authors thank Owen Arnold, Sean Giblin, Peter Leek, Connor Shelly and Brian Vlastakis for reviewing this manuscript and Martin Jackson and Oxford Instruments for useful discussions relating to this work.

\bibliographystyle{unsrtnat}
\bibliography{references} 
\newpage

\section{Supplemental}\label{Supplemental}

\subsection{Measurement setup}\label{msetup}
The dilution refrigerator used for this study is an \textit{Oxford instruments Proetox LX 850} \cite{OINS}, fitted with two \textit{Cryomech PT420} pulse tube cryocoolers each providing a quoted cooling capacity of 1.8 W at 4.2 K with 50 W at 45 K \cite{cryomech4K}. A platinum PT100 temperature sensor is installed at the PT1 stage (typical temperature 35-45 K), \textit{Cernox} temperature sensor at the PT2 stage (3-4 K) and RuOx sensors at the STL (0.7-1 K), CLD (80-150 mK) and mixing chamber stage (10-20 mK). Resistive cartridge heaters are thermally anchored on the PT1 stage (\SI{80}{\watt}), PT2 stage (\SI{80}{\watt}), STL (\SI{200}{\ohm}), CLD (\SI{100}{\ohm}) and MXC (\SI{100}{\ohm}). All sensor and heaters were pre-installed and integrated with a temperature controller by the manufacturer with the exception of the CLD heater which was added post-delivery using a pre-installed wiring loom and present for all measurements. 

\subsection{Measurement methodology}\label{method}
All measurements of system parameters are taken 1 week post system cool-down to base temperature to minimise time-dependent background sources of heat load, such as ortho-para He conversion, tunnelling effects in crystalline solids and thermal relaxation of materials. To determine the stabilisation time used for measurements of stage temperatures, pressures and flow rate, power is applied to each individual stage heater and the thermal time constant measured for each stage in the forward and reverse direction to account for any hysteresis. It is found that a 2 hr stabilisation time is required when changing the heat load configuration due to the long time constant associated with the thermalisation of the STL stage and subsequent modification to the circulation dynamics. 

The bases of the platform capacity map are the two-dimensional thermal load grids of the PT stages (PT1 and PT2 varied, other heaters off) and the DU (STL and MXC varied, other heaters off), which form the PT and DU maps respectively. These maps were measured in detail, with power steps on PT1, PT2, the STL and MXC of $\sim$\SI{4.8}{\watt}, $\sim$\SI{0.23}{\watt}, \qtyrange[range-units = single]{2}{20}{\milli\watt} and \qtyrange[range-units = single]{2}{5}{\micro\watt} respectively.

Additional combinations of heater powers were measured with sparse versions of the PT and DU maps. For these measurements, the maximum allowable range was reduced when pressure or temperature limits were encountered during combined loading scenarios. The sparse PT maps used \SI{10}{\watt} steps on PT1 and $\sim$\SI{1.5}{\watt} steps on PT2. For the DU map, this reduced density grid was made up of 0, 6, 20 and \SI{100}{\milli\watt} on the STL and 0, 4, 14, 30 and \SI{60}{\micro\watt} on the MXC. 

While the initial stage power limits were set to \SI{19.3}{\watt} for PT1, \SI{4.36}{\watt} for PT2, \SI{100}{\milli\watt} for the STL, \SI{32.5}{\milli\watt} for CLD and \SI{60}{\micro\watt} for the MXC based on user-defined typical or safe use, these were adjusted downward when specific power combinations led to system constraints being reached.

Capacity mapping measurements are performed programmatically, with recorded parameters being time averaged for 10 minutes after the 2 hr stabilisation time for every heater step. It should be noted that a slow temperature drift (\SI{1.8}{\milli\kelvin} per hour) was observed on PT1 during measurement cycles and corrected for, with other stages showing negligible drift (<1\% over measurement acquisition time). Post measurement of the platform capacity map the system was warmed up, opened and subsequently cooled down in the same configuration where we observed a <1\% of change in stage temperatures, with the exception of the mixing chamber which was approximately 2\% colder (\SI{7.28}{\milli\kelvin} compared with \SI{7.37}{\milli\kelvin}).

\subsection{Pressures and flow}\label{pressureandflow}
The circulation pressures and flow rate are key parameters extracted from system capacity mapping and as discussed in Section \ref{lpm} can be viewed as an important metric when factoring in system overheads and IO design. 
\begin{figure}[h]
\includegraphics[width=3.2in]{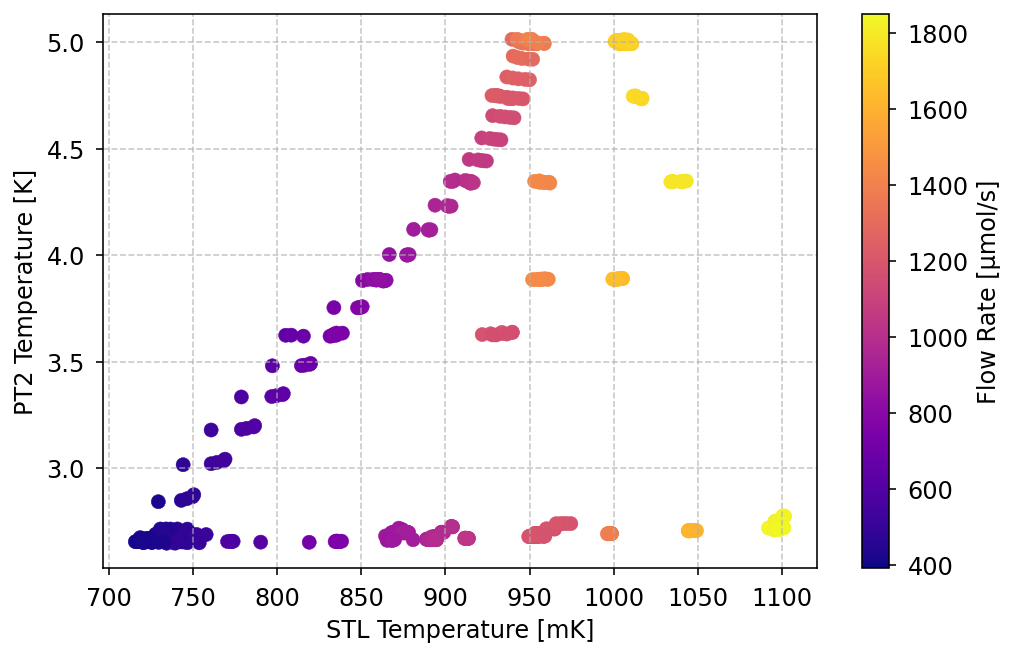}
\caption{\label{flow} 
Flow rate as a function of PT2 and STL stage temperate, extracted from the platform capacity map for a variety of loading configurations on the PT1, PT2, STL, CLD and MXC stages. Apparent discontinuity relates to displaying the data from very different loading scenarios on the same graph. 
}
\end{figure}
\begin{figure}
\includegraphics[width=3.2in]{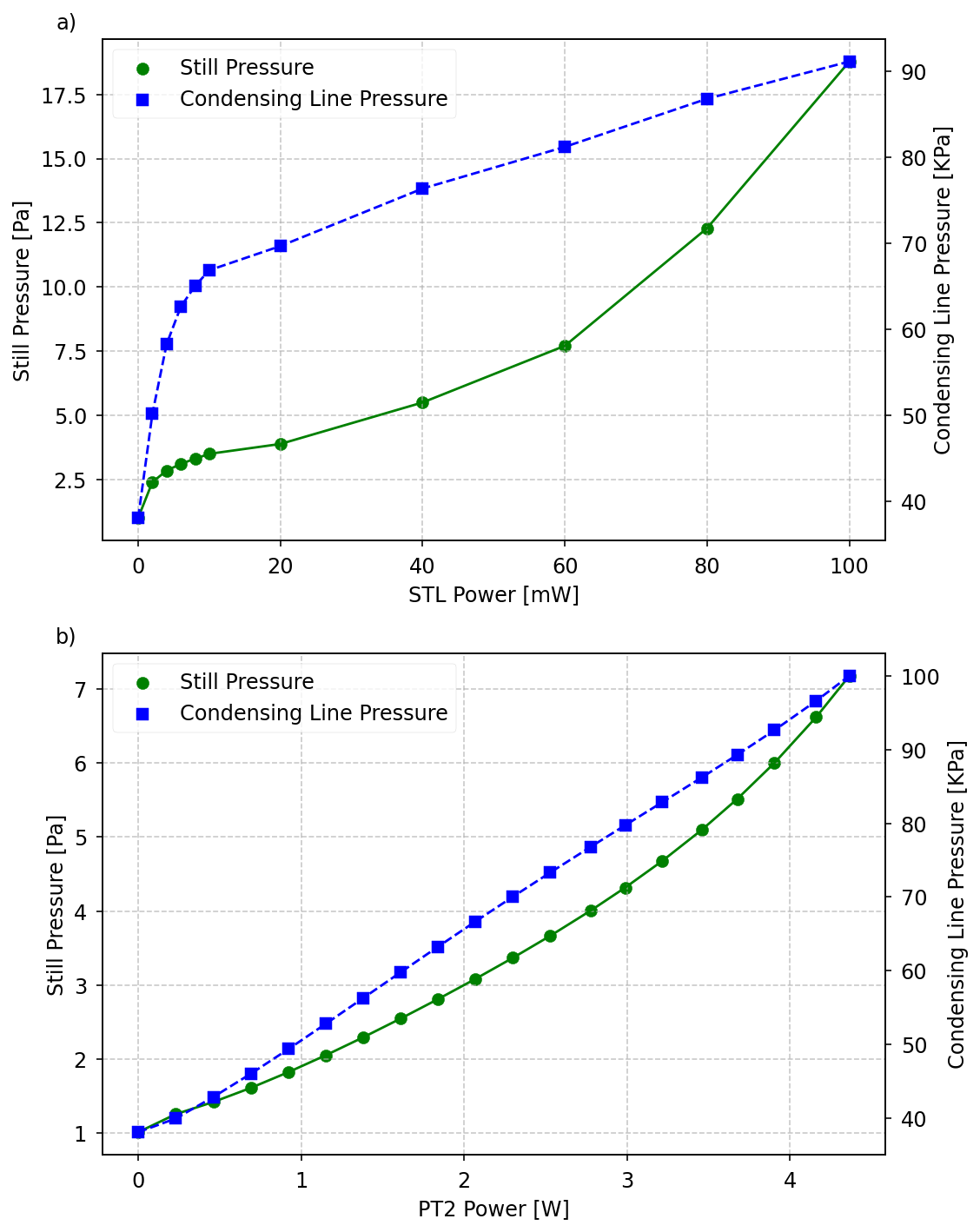}
\caption{\label{pressure} Condenser line and STL line pressures for A) STL loading B) PT2 loading.
}
\end{figure}

Loading the PT2 and STL stages have the greatest effect on the circulation dynamics, illustrated in Fig \ref{flow} and Fig. \ref{pressure}. STL loading resulting primarily in an increase in the STL line pressure due to increased evaporation and flow of the He gas in the circuit. This contrasts with PT2 loading, although it also results in an elevation of the STL temperature a key difference is that the increase in the stage temperature also results in a substantial increase of pressure in the condenser line. Simultaneous loading of both the STL and PT2 stage result in a combination of these effects further exacerbating the pressure increase in the circuit. 

\subsection{Validation payload modelling}\label{payloadcomp}
\begin{figure}
\centering
\includegraphics[width=2in]{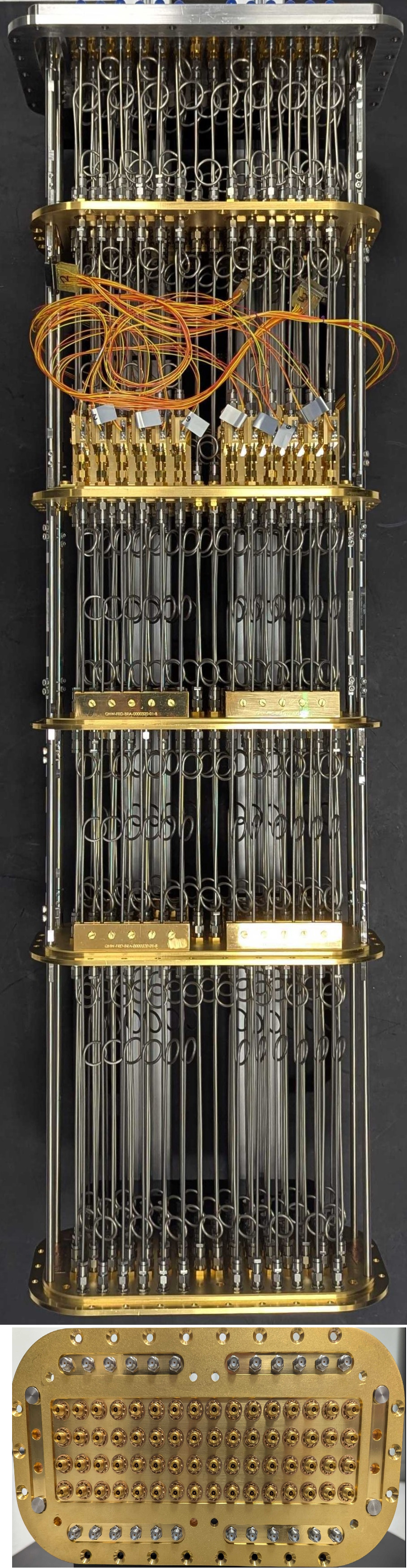}
\caption{\label{SIpic} The validation payloads is based around \textit{Oxford instruments Proteox} Secondary insert technology, with each stage mounting directly to the corresponding temperature stage of the dilution refrigerator. For handling and installation activities, support bars are attached to provide strain relief for the cabling assembly, once installed these support bars are removed such that the only connections between stages introduced is the cable assembly.}
\end{figure}
\begin{table*}[ht]
\begin{tabular}{c|c|cc|cc|}
\cline{3-6}
 \multicolumn{1}{c}{}&\multicolumn{1}{c|}{} &\multicolumn{2}{c|}{Manufacture data} & \multicolumn{2}{c|}{Material data} \\ \hline
\multicolumn{1}{|c|}{Stages Connected} & \multicolumn{1}{c|}{Length (m)}& \multicolumn{1}{c|}{Cable} & Load Per Wire & \multicolumn{1}{c|}{Cable} & Load Per Wire \\ \hline
\multicolumn{1}{|c|}{\multirow{2}{*}{Room Temp. - PT1}} & \multicolumn{1}{c|}{0.134}& \multicolumn{1}{c|}{SC219SS} & 25.13 mW & \multicolumn{1}{c|}{SS/PTFE} & 40.49 mW \\ \cline{2-6} 
\multicolumn{1}{|c|}{} & \multicolumn{1}{c|}{0.134}& \multicolumn{1}{c|}{SC219CN} & 15.8 mW & \multicolumn{1}{c|}{CuNi/PTFE} & 77.3 mW \\ \hline
\multicolumn{1}{|c|}{\multirow{2}{*}{PT1 - PT2}} & \multicolumn{1}{c|}{0.201}& \multicolumn{1}{c|}{SC219SS} & 343.03 µW & \multicolumn{1}{c|}{SS/PTFE} & 566.06 µW \\ \cline{2-6} 
\multicolumn{1}{|c|}{} & \multicolumn{1}{c|}{0.201}& \multicolumn{1}{c|}{SC219CN} & 552.08 µW & \multicolumn{1}{c|}{CuNi/PTFE} & 2.74 mW \\ \hline
\multicolumn{1}{|c|}{\multirow{2}{*}{PT2 - Still}}& \multicolumn{1}{c|}{0.163} & \multicolumn{1}{c|}{SC219SS} & 1.3 µW & \multicolumn{1}{c|}{SS/PTFE} & 3.09 µW \\ \cline{2-6} 
\multicolumn{1}{|c|}{} & \multicolumn{1}{c|}{0.173}& \multicolumn{1}{c|}{SC219NbTi}  & 697.08 nW & \multicolumn{1}{c|}{NbTi/PTFE} & 1.44 µW \\ \hline
\multicolumn{1}{|c|}{\multirow{2}{*}{Still - Cold Plate}}& \multicolumn{1}{c|}{0.163}& \multicolumn{1}{c|}{SC219SS} & 88.96 nW & \multicolumn{1}{c|}{SS/PTFE} & 241.11 nW \\ \cline{2-6} 
\multicolumn{1}{|c|}{} & \multicolumn{1}{c|}{0.171}& \multicolumn{1}{c|}{SC219NbTi} & 24.6 nW & \multicolumn{1}{c|}{NbTi/PTFE} & 101.97 nW \\ \hline
\multicolumn{1}{|c|}{\multirow{2}{*}{Cold Plate - MXC}}& \multicolumn{1}{c|}{0.236} & \multicolumn{1}{c|}{SC219SS} & 0.83 nW & \multicolumn{1}{c|}{SS/PTFE} & 2.26 nW \\ \cline{2-6} 
\multicolumn{1}{|c|}{} & \multicolumn{1}{c|}{0.246}& \multicolumn{1}{c|}{SC219NbTi} & 0.21 nW & \multicolumn{1}{c|}{NbTi/PTFE} & 0.96 nW \\ \hline
\end{tabular}%
\caption{\label{cable_loads_table} Per cable heat load estimations for the cabling populating the validation payload. Manufacture data uses the thermal conductivity for the individual cable types specified in \protect\cite{Coco}. Material data uses the thermal conductivity specified in \protect\cite{NIST} and \protect\cite{1501.07100} for the constituent materials used to construct the cable, given the material and mechanical specifications provided by the manufacture.}
\end{table*}

A photograph of the validation payload is shown in Fig. \ref{SIpic}. The payload has a total of 64 input lines and 24 output lines, configured as described in Fig. \ref{DR Cart}b). The total heat load on each stage introduced by the payload can be calculated by summing the contributions from the passive heat load owing to stage interconnections and active sources from dissipative componentry such as amplifiers, or power dissipated in resistive elements due to applied component biases and measurement signals. In the case of Section \ref{pmodel}, only the effect of the passive loading from the addition of interstage connections is considered (LNAs unpowered with no application of measurement signals or biases). 

\begin{figure}
\includegraphics[width=3.2in]{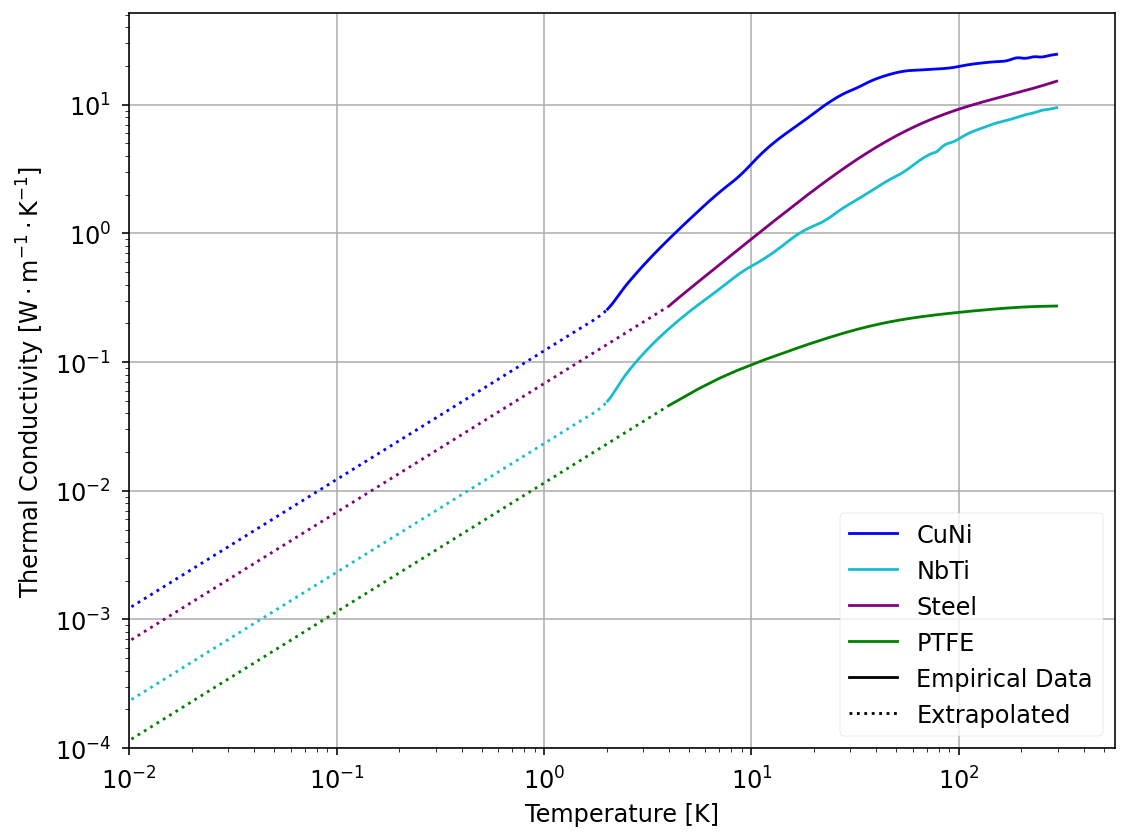}
\caption{\label{conductivities} Material thermal conductivities as a function of temperature. Empirical data reference data is represented by a solid line, with the linear extrapolation to 0 K shown as a dotted line.
}
\end{figure}

To calculate the associated passive heat load presented by an individual interconnection we model the static heat load $Q$  using the integral form of Fourier's law:
\begin{eqnarray}
Q = \frac{A}{L}\int_{T_{C}}^{T_{H}}k(T)dT
\label{eq:three},
\end{eqnarray}
where $A$ is the cross-sectional area of the interconnection, $L$ is its length, $k(T)$ is the temperature dependent thermal conductivity and $T_{H}$ and $T_{C}$ are the temperatures of the hot and cold ends, respectively. Mixed material interconnects, such as coaxial cables with an inner conductor, dielectric and outer conductor, are decomposed and treated as parallel single material thermal links with an effective cross sectional area of specified length. Temperature dependent thermal conductivity can be estimated using empirical data from \cite{NIST} and \cite{1501.07100} to provide fixed point reference data and appropriate theoretical models. The $k(T)$ data for the materials used in the validation payload cabling, over the relevant operational temperature ranges, is shown in Fig. \ref{conductivities}. A linear extrapolation to 0 K is used in the absence of data at milli-Kelvin temperatures consistent with the approach taken in \cite{1501.07100}.

An alternative source of thermal conductivity data can be obtained by referencing manufacture provided data-sheets for the individual cables (\textit{Coax Co.} \cite{Coco}), again using a linear extrapolation to 0 K to extend the data to lower temperatures. However, here we find disagreement between the manufacturer values and the modelled values based on the aforementioned references (\cite{NIST} and \cite{1501.07100}), leading to a range of heat loads attributed to the cabling depending on which reference source is used.  

As an example of the uncertainty of stage interconnect thermal coupling, different coupling configurations for the decomposed NbTi coaxial cabling between PT2 and the MXC are explored. This is motivated by a lack of 0 dB attenuation of the output lines at the intervening stages (STL and CLD), which has otherwise formed the basis of the assumption of thermalisation of the decomposed constituents and the stages. Arrangements where the inner conductor, dielectric and outer conductor are either coupled to each stage or only between PT2 and the MXC were modelled, with significant differences in passive heat load noted between each arrangement.

A comparison of thermal loads per wire calculated from manufacturer data and thermal conductivities derived directly from material reference values, with cables and cable constituents coupled at each stage, is presented in Table \ref{cable_loads_table}. We see that the largest values are produced where the thermal conductivity contribution is entirely derived from material reference values. These discrepancies, and the lack of agreement with the inferred heat loads in Table \ref{vladtable}, highlight a gap in the available low temperature thermal conductivity reference data and current methods for modelling thermal conductivities and payload thermal loads.

Once a total distributed heat load has been calculated for a given payload this can be input to the platform capacity map and the modified stage temperatures can be estimated. This in turn modifies the passive heat load generated from interstage connections, which are recalculated and the new distributed heat loads input to the platform capacity map, with this process iterating until the convergence is reached, representing thermal equilibrium. At this point, the platform metrics, including inferred stage temperatures, for the given payload can be determined. 
\end{document}